\documentclass[aps,pra,superscriptaddress,twocolumn,amsmath,amsfonts,amssymb,floatfix]{revtex4-1}
\usepackage{enumerate}
\usepackage{mathtools}
\usepackage{amsmath}
\usepackage{graphicx}
\usepackage{epsfig}
\usepackage{sidecap}
\usepackage{hyperref}
\usepackage{color}
\usepackage{subfigure,array}
\usepackage{verbatim}

\usepackage{enumerate}
\usepackage{bm}

\begin{document}

\title{Investigation of Quantum Droplet: An Analytical Approach}

\author{Argha Debnath}

\author{Ayan Khan} \thanks{ayan.khan@bennett.edu.in}
\affiliation{Department of Physics, School of Engineering and Applied Sciences, Bennett University, Greater Noida, UP-201310, India}

\begin{abstract}
Recent observations of droplets in dipolar and binary Bose-Einstein condensate (BEC) motivates us to study the theory of droplet formation in detail. Precisely, we are interested in investigating the possibility of droplet formation in a quasi-one-dimensional geometry. The recent observations have concluded that the droplets are stabilized by the competition between effective mean-field and beyond mean-field interaction. Hence, it is possible to map the effective equation of motion to a cubic-quartic nonlinear Schr\"odinger equation (CQNLSE). We obtain two analytical solutions of the modified Gross-Pitaevskii equation or CQNLSE and verified them numerically. Based on their stability we investigate the parameter regime for which droplets can form. The effective potential allows us to conclude about the regions of soliton domination and self-bound droplet formations.
\end{abstract}

\maketitle

\section{Introduction}
The collective behavior of particles at ultra-low temperature is a fascinating topic ever since the experimental observation of atomic Bose-Einstein condensate (BEC) \cite{anderson,hulet,ketterle}. 
The experimental success had raised the curtain for new domains of research \cite{stringari1999,stringari2008,bloch}. Over the time, the experimentalists have achieved greater control over the atomic alkali gases by means of magneto-optic setup. Moreover,
they can tune the atom-atom interaction via Feshbach resonance which effectively means, changing of $s$-wave scattering length by tuning an external magnetic field. These unique features have enabled multi-facet research in ultra-cold atomic gases \cite{grimm}. 

Very recently, a unique liquid-like state in a BEC mixture \cite{cabrera1} has been reported. This bizarre new state demands serious attention because the prevailing conception of the liquid state is heavily influenced by the theory of Van der Waals. 
However, these newly emerged droplets in ultra-cold and extremely dilute atomic gases do not explicitly follow the common theoretical perception as predicted by van der Waals \cite{barbut2}. These are purely quantum mechanical in nature and manifestation of quantum fluctuations \cite{barbut1,barbut3}.
These droplets are small clusters of atoms self-bound by the interplay of attractive and repulsive forces. The origin of the attractive force can be modeled in the purview of standard mean-field (MF) theory whereas the repulsive force originates from the beyond mean-field correction \cite{sala1}. The underlying theory relies on the Lee-Huang-Yang's (LHY) correction \cite{lee} to the mean-field Gross-Pitaevskii (GP) equation \cite{gross,pitae}. In a binary BEC, the mean-field and LHY term depend on the balance of inter and intraspecies coupling constants. Even before the experiment, it was proposed theoretically that, if the square of the interspecies coupling is greater than the product of the intraspecies coupling then the collapse of the binary mixture is suppressed and a dilute liquid-like droplet state emerges \cite{petrov}. The emergence of this phase has opened several new avenues as these droplets describe truly many-body quantum effect.

The current framework was first proposed while discussing the possibility of collapse due to attractive interaction in Bose-Bose mixture \cite{petrov}. However, the first experimental observation was on dipolar condensate of $^{164}Dy$ \cite{barbut1,barbut3} and subsequently the theoretical description came to light \cite{santos}. Later the droplets were observed for a mixture of two hyperfine states in $^ {39} K$ \cite{cabrera1,fattori}. This was followed by an observation of transition from bright solution to quantum droplets \cite{cabrera2}. Quantum droplet is also observed in a heteronuclear bosonic mixture of $^{41}K$ and $^{87}Rb$ \cite{fort}. Of late a self-consistent derivation of modified GP equation where LHY correction is incorporated through quantum fluctuation has been proposed \cite{sala2}. We have also noted significant theoretical description of the collective modes across the soliton-droplet cross over \cite{sala1}, existence of vortex quantum droplets \cite{malomed}, dynamics of purely one-dimensional droplet \cite{astra} and its collective excitations \cite{tylutki}. 

Here, we plan to analytically analyze the two component BEC in quasi-one-dimensional (Q1D) system as described in Fig.~\ref{cartoon}. At this juncture, it is worth mentioning that a numerical investigation of quantum liquid in for dipolar BEC in Q1D geometry has very recently been reported \cite{edmonds}. Our focus also gels well with the current interests on droplets at lower dimension which includes, a comprehensive analysis on the role of LHY term in suppressing the collapse in quasi two-dimensional system \cite{chen}. Of late, a possible connection has also been drawn between the droplets and modulational instability in one dimension \cite{khare}.

The formulation of the current problem follows the prescription of Ref.\cite{cabrera1}, where the mixture of two hyper-fine states of $^{39}K$ was studied assuming that both the components occupy the same spatial mode. This ensures the two-component nonlinear Schrodinger equation is reduced to an effective one component equation. Then we reduce the 3+1-dimensional problem to 1+1-dimensional problem following the prescription of Ref.\cite{atre}. The quasi 1-D system now consists of two nonlinear terms where the cubic term defines the effective mean-field (EMF) two-body interaction, and the quartic term is the signature of beyond mean-field or LHY contribution. Our primary goal is to find out an analytical solution for this quasi 1-D cubic-quartic nonlinear Schr\"odinger equation (CQNLSE). The next objective is to explore the droplet state. Here we note that, very recently we have proposed cnoidal solutions of CQNLSE \cite{argha}. There, we have used a cnoidal potential to stabilize the analytical solutions.

In this piece of work, we specifically focus on (i) obtaining an analytical solution of quasi 1D CQNLSE; (ii) validation of the analytical solution through numerical calculation via split-step Crank-Nicolson (CN) method; (iii) investigation of the stability for the obtained solutions via Vakhitov-Kolokolov (VK) criterion \cite{vakhitov1973stationary}; (iv) linear stability analysis which includes investigation of modulational instability \cite{azbel} and calculation of the growth rate corresponding to the perturbation eigenmodes \cite{soto}; (v) examination of the the droplet region using the analytical solution where we calculate the equilibrium density and critical density of the liquid-like state. The equilibrium density is noted as the transition point between bright soliton-like state to stable liquid like state. The critical density, beyond which the droplets disappear, is calculated and corroborated with the theory. We also demonstrate the existence of a density plateau for higher number of particles signifying the creation of quantum liquid.

It must be noted that the dynamics of purely 1D droplet was quite extensively studied in Ref.\cite{astra}, however the current investigation lies in the realm of quasi-one-dimensional systems which can be more amenable experimentally. It is well known fact that condensate formation is not possible in 1D, therefore it is common to study the condensate formation in a quasi 1D geometry where the Bose gas is allowed to expand in an optical waveguide which has enabled us to observe exotic structures like the bright soliton trains \cite{salomon,strecker}. However, it is possible to correctly predict the energy of a weakly interacting Bose gas using Bogoliubov theory which assumes the existence of condensate in one dimension \cite{lieb1,popov,petrov1}. Here, by stating a quasi 1D system, we assume that $\sqrt{na^3}<1$ however $\frac{1}{|\sqrt{n_{1D}a_{1D}}|}>1$ \cite{petrov2}. $n$ and $a$ stand for the particle density and $s$-wave scattering length whereas $n_{1D}$ and $a_{1D}$ are the density and scattering length respectively in one dimension. It can be noted that $a_{1D}=2\hbar^2a/a_{\perp}^2m$, with $a_{\perp}$ being the characteristic length scale of the trap and $m$ being the mass of the particle \cite{rejish}. Nevertheless, our objective is to remain more towards the right side of the dimensional crossover whereas a strict 1D system is more towards the left of the crossover where $\frac{1}{|\sqrt{n_{1D}a_{1D}}|}<1$ as suggested in Ref.~\cite{petrov2}. Mathematically, a quasi-one-dimensional system leads to a CQNLSE whereas a 1D system can be described by a quadratic-cubic NLSE \cite{petrov1}.


In this paper, we report our results in the following sequence, in Sec.\ref{model}, we elaborate the theoretical model corresponding to the binary condensate, it's mapping to one component extended Gross-Pitaevskii (GP) equation and dimensional reduction of the system from 3+1 to 1+1 dimension. We obtain the analytical solutions corresponding to the extended GP equation and we analyze the stability of the solutions in Sec.\ref{sol}. The possibility of droplet formation is explicated in Sec\ref{drop}. We draw our conclusion in Sec.\ref{con}.

\section{Theoretical Model}\label{model}
\begin{figure}
\includegraphics[scale=0.25]{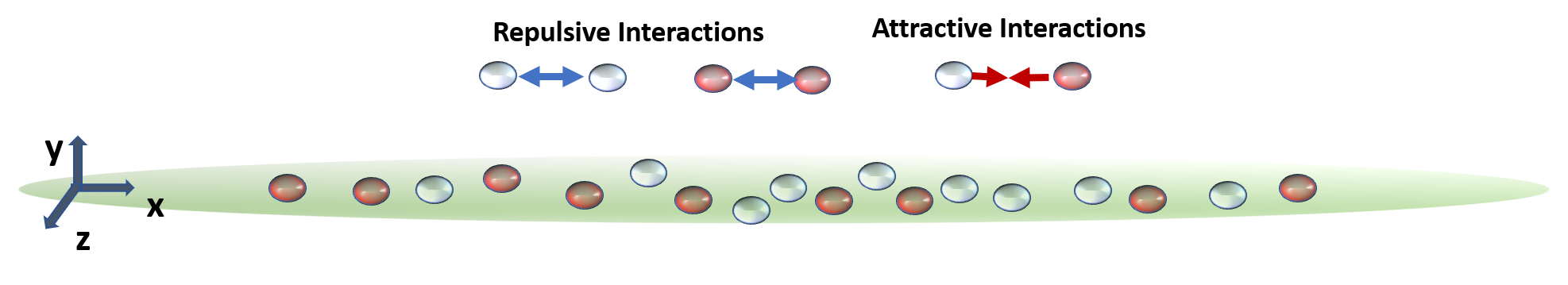}
\caption{(Color online) Schematic representation of two-component BEC in a quasi-one-dimensional confinement. The bluish and reddish spheres present atoms in the two different hyper-fine states. The intraspecies and interspecies interactions are repulsive and attractive respectively.}\label{cartoon}
\end{figure}
Here, we consider homonuclear bosonic mixture similar to that of Ref.\cite{cabrera1}, where two hyperfine states of $^{39}K$ took part in the experiment, in a Q1D geometry. 
The situation can be visualized via Fig.~\ref{cartoon} where the blueish and reddish spheres represent the different species of atoms (atoms in two hyperfine states) 
distributed in an effectively one-dimensional cigar shaped trap. The intraspecies interactions ($a_{11}$ and $a_{22}$) are repulsive in nature and the interspecies interaction ($a_{12}$ or $a_{21}$) is attractive. 
The experimentally observed droplets are small clusters of atoms, self-bound by the balance between the attractive and repulsive forces. 
In binary BEC based on the strength of intraspecies and interspecies interactions, it possible to define three distinct ground states. If $a_{11}, a_{22}$ and $a_{12}$ all are repulsive then one expects a transition
between the miscible and the immiscible phase. However, the mixture can also collapse if the interspecies interaction is negative enough to counter the repulsive intraspecies interactions. It can be shown that if $a_{12}>\sqrt{a_{11}a_{22}}$ then the mixture is in immiscible phase, if $-\sqrt{a_{11}a_{22}}<a_{12}<\sqrt{a_{11}a_{22}}$ then the condensate is in miscible phase and when $-\sqrt{a_{11}a_{22}}>a_{12}$ then the condensate collapses. Now, it is possible to model the EMF interaction strength $\delta a\propto(a_{12}+\sqrt{a_{11}a_{22}})$ which is close to the collapsing regime. If $\delta a\lesssim0$ then the beyond mean-field (BMF) contribution becomes significant. Let us define the BMF contribution as $\delta a'\propto(\sqrt{a_{11}a_{22}})^{5/2}$ \cite{cabrera2}. In the miscible phase and close to the collapse point, we
can describe the system with an effective single component GP equation by neglecting the spin excitations. This criterion can be full filled by considering the two components occupy the same spatial mode. The resulting
one component equation of motion can be defined as \cite{cabrera1},
\begin{eqnarray}
i\hbar\frac{\partial\Psi}{\partial t}&=&\left[\left(-\frac{\hbar^2}{2m}\nabla^2+V_{trap}\right)+U|\Psi|^2+U'|\Psi|^3\right]\Psi,\nonumber\\
&&\label{3dbgp}
\end{eqnarray}
where, $U=\frac{4\pi\hbar}{m}\delta a$, $U'=\frac{256\sqrt{\pi}\hbar^2\delta a'}{15m}$ and $m$ being the mass of the atoms. 
The equation is quite unique as there exist two types of nonlinearity, the usual cubic nonlinearity as well as an additional quartic nonlinearity.
Here, it can be noted that nonlinear Schr\"odinger equation with cubic and quintic nonlinearity, ($\propto|\Psi|^4\Psi$) is quite common in nonlinear optics \cite{peleg} and BEC \cite{paredes}.
However, it is not the same for quartic nonlinearity ($\propto|\Psi|^3\Psi$).
At this juncture, we also like to note that, in the early days of 21$^{st}$ century, the possibility of droplet formation was explored via quintic nonlinearity as well \cite{bulgac}.
Nevertheless, the repulsive term possessing an unusual quartic
dependence manifests the beyond mean-field contribution, which is not well studied till date \cite{argha}.
Therefore, we are primarily motivated to obtain an analytical solution for a NLSE which has both cubic and quartic nonlinearity.

Further, in Eq.(\ref{3dbgp}), 
$V_{trap}$ describes the external potential. It is possible to describe the external potential in terms of the transverse component ($V_T(y,z)=\frac{1}{2} m \omega_{\perp}^2(y^2+z^2)$) and longitudinal component ($V_L(x)$). Here, $\omega_{\perp}$ is the transverse trap frequency. The potential along the longitudinal direction is defined as, $V_L(x) = \frac{1}{2} m \omega_{0}^2 x^2$ with $\omega_{0}$ being the longitudinal trap frequency. In cigar-shaped BEC the transverse trapping frequency ($\omega_{\perp}$) is typically more than $10$ times
the longitudinal frequency ($\omega_{0}$). It can be noted that in the early days of ultra-cold atom research, soliton trains were observed in an one dimensional optical waveguide whoes longitudinal ($\omega_x$) and transverse trap frequencies ($\omega_{\perp}$) were set at $2\pi\times 50$Hz and $2\pi\times710$Hz respectively \cite{salomon}. Since, the characteristic length scale happens to be $a_{\perp}=\sqrt{\frac{\hbar}{m\omega_{\perp}}}$, therefore in a quasi-one-dimensional geometry $a_x/a_{\perp}\sim\sqrt{10}$. Here, $a_x$ is noted as $a_x=\sqrt{\frac{\hbar}{m\omega_{0}}}$ \cite{parola}. 
This implies that the interaction energy of the atoms is much less than the kinetic energy in the transverse direction. 

Consequently, it is possible to reduce Eq.(\ref{3dbgp}) to an effective one-dimensional equation. In order to perform the dimensional reduction, we have made use of the following ansatz, 
\begin{eqnarray}
\Psi(\mathbf{r}, t) &=& \frac{1}{\sqrt{2\pi a_B}a_{\perp}}\psi\left(\frac{x}{a_{\perp}},\omega_{\perp}t\right) e^{\left(-i\omega_{\perp}t-\frac{y^2+z^2}{2a_{\perp}^2}\right)},\nonumber\\\label{ansatz1}
\end{eqnarray}
where, $a_B$ is Bohr radius.

Applying the ansatz from Eq.(\ref{ansatz1}) in Eq.(\ref{3dbgp}) we obtain the quasi-one-dimensional (cigar-shaped) extended GP equation as noted below,
\begin{widetext}\begin{eqnarray}
i\frac{\partial\psi(x,t)}{\partial t} = & & \left[ - \frac{1}{2}\frac{\partial^2}{\partial x^2} + \frac{1}{2} K x^{2}+\tilde{g} |\psi(x,t)|^2+\tilde{g}' |\psi(x,t)|^3 \right]\psi(x,t)\label{bgp},
\end{eqnarray}\end{widetext}
where, $\tilde{g}=2\delta a/a_B$, 
$\tilde{g}'=(64\sqrt{2}/15\pi)\delta a'/(a_B^{3/2}a_{\perp})$ and $K = \omega^{2}_{0}/\omega^{2}_{\perp}$. 
Here, it is important to note that $x$ and $t$ are now actually dimensionless, i.e. $x\equiv x/a_{\perp}$ and $t\equiv\omega_{\perp}t$. From here onward, we will follow this dimensionless notation of $x$ and $t$. 


In this article, our main focus is to explicate the interplay between EMF and BMF interactions hence we exclude the effect of harmonic confinement and the system can become quasi-homogeneous. Experimentally the system can be reduced to a quasi-homogeneous setup by considering transverse confinement is much stronger compared to the longitudinal confinement ($\omega_{0}<<\omega_{\perp}$) resulting $K\rightarrow0$. The next objective is to obtain analytical solution for Eq.(\ref{bgp}) assuming $K=0$. 

\section{Solutions}\label{sol}
In this section, we elaborate on the mathematical scheme to derive the analytical solution for the extended GP equation and analyze the stability of the obtained solutions. To start with, we write the wave function such a way that, $\psi(x,t)=\rho(x,t)\exp{\left[i\left(\chi(x,t)+\mu_0 t\right)\right]}$ where $\rho(x,t)$ leads to the amplitude contribution and $\chi(x,t)$ is the non-trivial phase, $\mu_0$ being the chemical potential. Applying this ansatz in Eq.(\ref{bgp}) we yield two equations, namely imaginary and real equation respectively such that,
\begin{eqnarray}
\rho_t&=&-\chi_x\rho_x-\frac{1}{2}\chi_{xx}\rho\label{imaginary}\\
-\chi_t\rho&=&-\frac{1}{2}\left(\rho_{xx}-\chi_x^2\rho\right)+\tilde{g}\rho^3+\tilde{g}'\rho^4+\mu_0\rho.\label{real}
\end{eqnarray} 
Eq.(\ref{imaginary}) leads to the continuity equation and if we transform the equation in center of mass frame, i.e., $\zeta=x-ut$, then we obtain,
\begin{eqnarray}
\chi_{\zeta}=u+\frac{C_0}{\rho^2}.\label{chi}
\end{eqnarray}
Here, $u$ defines the velocity of the wave and $C_0$ is the integration constant.
Eq.(\ref{real}) in the comoving frame can be rewritten as,
\begin{eqnarray}
\chi_{\zeta}u\rho=-\frac{1}{2}\left(\rho_{\zeta\zeta}-\chi_{\zeta}^2\rho\right)+\tilde{g}\rho^3+\tilde{g}'\rho^4+\mu_0\rho.\label{rho}
\end{eqnarray}
Applying Eq.(\ref{chi}) in Eq.(\ref{rho}) we obtain,
\begin{eqnarray}
\rho_{\zeta\zeta}+(u^2-2\mu_0)\rho-2\tilde{g}\rho^3-2\tilde{g}'\rho^4&=&0\nonumber\\
\textrm{or,}\,\,\frac{d^2\rho}{d\zeta^2}+\left(g\rho^2-g'\rho^3+2\gamma\right)\rho&=&0.\label{cq1}
\end{eqnarray}
To derive Eq.(\ref{cq1}) it is important to consider that the phase and amplitude are uncorrelated which allows us to set $C_0=0$ \cite{khan4}. We also note that $\gamma=u^2/2-\mu_0$.
Further, we assume
$g=-2\tilde{g}$ and $g'=2\tilde{g}'$ implying a two-body effective mean-field interaction is attractive and LHY contribution is repulsive. The minimum criterion for droplet formation is that these two interactions
must be competing. Otherwise, we will not be able to see any qualitative change in the behavior of the system.

We consider an ansatz solution such that,
\begin{eqnarray}
\rho(\zeta)=\frac{A}{1+\sqrt{1-A}\cosh({\sqrt{\xi}\zeta)}},\label{ansatz}
\end{eqnarray}
where $\sqrt{\xi}$ is the inverse of coherence length. Applying the ansatz in Eq.(\ref{cq1}) we obtain a set of condition for which Eq.(\ref{ansatz}) is a solution of Eq.(\ref{cq1}).
The constrained conditions read,
\begin{eqnarray}
A&=&\frac{-\xi\pm\sqrt{-6\gamma g+3g\xi+\xi^2}}{g},\nonumber\\
\xi&=&6\gamma+g,\nonumber\\
\gamma&=&\frac{g'-g}{2},\nonumber\\
|g|&=&2g',\,\textrm{or}\,|g|=3g'.\label{cond}
\end{eqnarray}

The last equation in Eq.(\ref{cond}) implies that it is possible to obtain an analytical solution if and only if the beyond mean-field interaction is half or one third (i.e.,$g'=|g|/2$ or $|g|/3$) of
the effective mean-field interaction and repulsive in nature. Hence, we can write the solutions as, 
\begin{eqnarray}
\rho(\zeta)&=&\frac{12\mu_g}{1+\sqrt{1-12\mu_g}\cosh{\left(\sqrt{\frac{g}{2}}\zeta\right)}}\,\textrm{for}\, |g|=2g'\label{con1}\\
&=&\frac{1+12\mu_g}{1+\sqrt{12\mu_g}\cosh{\left(\sqrt{g}\zeta\right)}}\,\textrm{for}\, |g|=3g'. \label{con2}
\end{eqnarray}
Here, $\mu_g=\mu_0/g$. Using the constrained conditions, we can also evaluate $\xi$ which is actually related to the two-body interaction 
via $\xi=-|g|/2$ or $-|g|$. This implies that the localized structures can only sustain if and only if $g<0$ or the effective mean-field interaction is attractive.
We must note here that for real solution, $\mu_g>0$, $\mu_0<0$ and correspondingly $\gamma>0$. In the subsequent discussions we will use $|g|=1$ for uniformity. The effect of variation of $|g|$ can be a matter of future interest.

Here, our major objective is to understand the interplay between EMF and BMF interaction for the formation of droplets and the role of chemical potential. Hence, we define the relationship between normalization $N$ and
chemical potential $\mu_0$ as,
\begin{eqnarray}
N=\left\{\begin{array}{c}
\sqrt{\frac{2}{g}}\left[\sqrt{\mu_I}\ln\left[\frac{2\sqrt{\mu_I}}{\sqrt{\mu_I}-1}-1\right]-2\mu_I\right]\\
\frac{(1+\mu_I)^2}{\sqrt{g}(1-\mu_I)}\left[\ln\left[1-\frac{2}{\mu_I}\left(\sqrt{1-\mu_I}+1\right)\right]-2\right]\end{array}\right.\nonumber\\\label{N}
\end{eqnarray}
Here, the first equation derived from Eq.(\ref{con1}) (assuming $\mu_I<1$). Likewise, $N$ is again calculated from Eq.(\ref{con2}) and noted in the second equation. 
We also recall that $\mu_g=\mu_0/g$ and for the convenience of calculation we have denoted, $12\mu_g=\mu_I$. 
$N$ can also be noted as the number of particles associated with the formation of localized wave and scaled by $N_0$ where $N_0$ defines the particle number obtained from the constant background density solution such that $N_0=2a_B\left(\frac{15\pi a_{\perp}}{64}\right)^2\left(\frac{\delta a}{\delta a'}\right)^2$.
\begin{figure}
\includegraphics[scale=0.3]{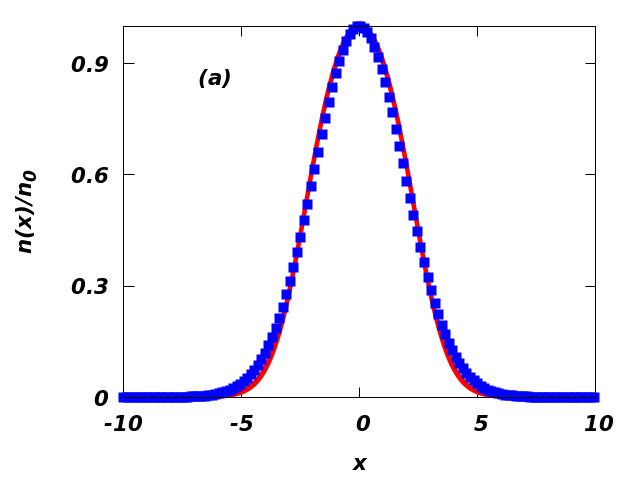}
\includegraphics[scale=0.3]{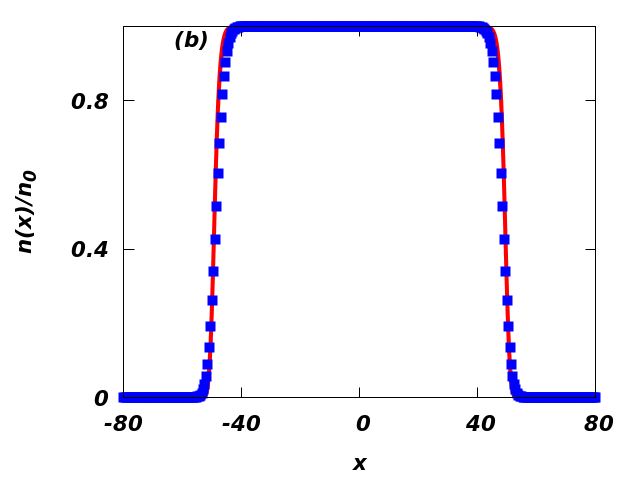}
\caption{(Color online) The figures described the comparison between the obtained analytical solution using Eq.(\ref{con2}) and numerical solution of Eq.(\ref{bgp}). The solid red line described the analytical result, and the blue squares represents numerically obtained solution. (a) depicts the low particle number solitonic regime (figure created for $N=4$) and (b) described high particle number ($N=100$) droplet regime. The density is normalized by $n_0$ where $n_0$ is $n(x)|_{x=0}$.}\label{nu_an}
\end{figure}

\subsubsection*{Numerical Analysis}
Now, we corroborate our analytical result with numerical simulation. For this purpose, split-step Crank-Nicolson (CN) method with imaginary time propagation is quite useful. It is well accepted that, for stationary ground states, imaginary-time propagation method is very accurate, and convergence is quite fast. This method also happens to be very robust. Hence, we employ the CN algorithm following Ref.\cite{muru} for our model. In Fig.~\ref{nu_an}, we compare the analytical and numerical result where the solid red line is our analytical solution from Eq.\ref{con2} and the blue solid squares are the numerically obtained solution of Eq.(\ref{bgp}). Fig.~\ref{nu_an}(a) and (b) corresponds to bright soliton like state ($N=4$) and liquid like state ($N=100$) respectively. 

Since, our analytical solution is constrained through a relationship between MF and BMF interaction strength ($|g|=2g'$ and $3g'$) thus we can use our numerical result for variety of interaction parameters to study beyond mean-field phenomena more closely. We did check our numerical solution for $\kappa=g'/|g|=0.1$, $0.5$ and $0.8$ starting with a seed solution such as, $\textrm{sech}(x)$. The results are in accordance with our understanding of the role played by BMF interaction such as progressive flattening of density with increasing BMF interaction strength. However, in this article we restrict ourselves as it goes beyond the purview of the current objective.

\subsubsection*{Stability Analysis}
Before proceeding to any discussion related to droplet formation, it is important that we evaluate the stability of the obtained solutions. For this purpose, we intend to employ the well-known VK (VK) criterion \cite{vakhitov1973stationary}. The approach we use is based on a modification of the soliton perturbation theory \cite{kivshar1989} under the condition of slow (almost adiabatic) evolution of solitons near the instability threshold. The VK criterion has been widely used in determining the stability of the solutions of nonlinear Schr\"odinger equation (NLSE), which predicts the parameter regime in chemical potential where the soliton’s amplitude can grow or decay exponentially \cite{khan7}. The VK criterion states that a necessary stability condition is a positive slope in the dependence of the number of atoms on the chemical potential. If, $\mathcal{N_{\mu}}> 0$, the solution is found to be stable and for $\mathcal{N_{\mu}} < 0$, the solution is unstable. One must note that the condition $\mathcal{N_{\mu}}= 0$ provides the instability threshold (TH) where, $\mu = \mu_{TH}$ \cite{pelinovsky1996,sakaguchi2010,khan7}. Here $\mathcal{N_{\mu}}=\frac{\partial N}{\partial \mu}$.

\begin{figure}
\includegraphics[scale=0.25]{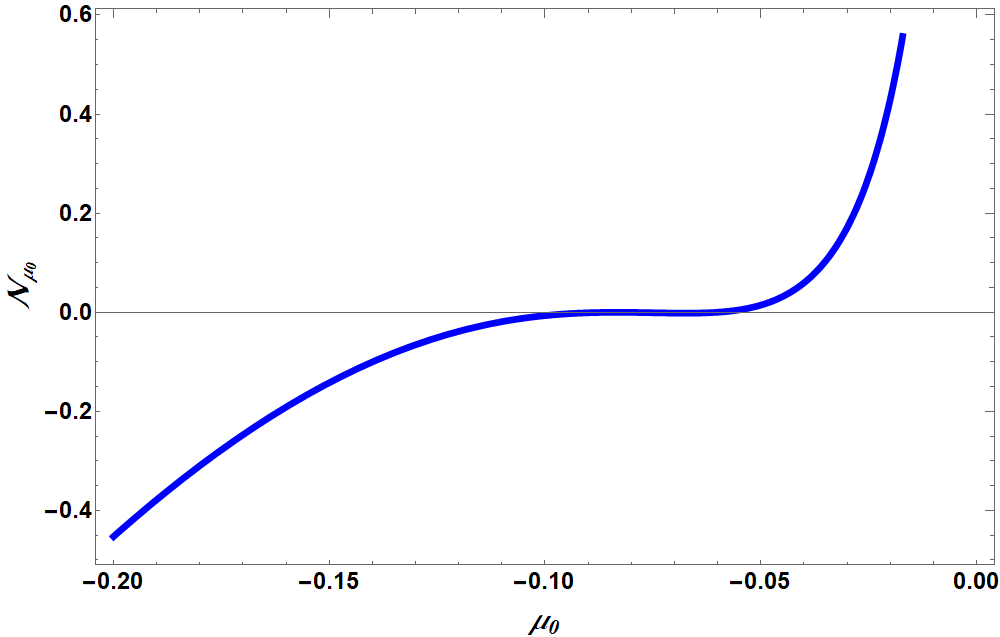}
\caption{(Color online) The stability criterion is inspected for the solution described in Eq.(\ref{con2}) which depicts a zero crossing at $\mu_0=-0.06$ for $|g|=1$.}\label{stability}
\end{figure}

In order to obtain the stability criterion of the given solutions, we calculate $\mathcal{N}_{\mu_0}$ from Eq.(\ref{con1}) as well as Eq.(\ref{con2}) as $\mathcal{N}_{\mu_0}=\frac{\partial N}{\partial \mu_0}$
A primary inspection leads to the conclusion that the first case or Eq.(\ref{con1}) does not lead to any stable solution. However, it's possible to obtain a region where $\mathcal{N}_{\mu_0}$ is positive thereby suggesting a stable solution regime from the second solution or Eq.(\ref{con2}). The behavior of $\mathcal{N}_{\mu_0}$ is noted in Fig.~\ref{stability}. We observe that the threshold value is $-0.06$ after which $\mathcal{N}_{\mu_0}>0$. However, $\mathcal{N}_{\mu_0}$ diverges as $\mu_0\rightarrow0$. We are unable to find any region of stability for positive $\mu_0$. 

Next, we perform the linear stability analysis of the second solution. In the realm of linear stability, our first objective is to calculate the modulational instability (MI) where the applied perturbation is considered as plane waves. It must be noted that MI plays a crucial role in nonlinear systems and quite recently through a remarkable experiment it was shown that soliton trains were created by the MI \cite{luo}. Very recently, the role of MI has been discussed in the context of droplet formation in a purely 1D system \cite{khare}. Motivated by these recent developments, we perform the MI analysis which reveals the region of instabilities in the parameter space.

Here, we apply a small perturbation to the stationary solution
such that, $\psi(x,t)=\psi_0(x)+\delta\psi(x,t)$ provided, $\delta\psi<<1$. Now, substituting the solution in Eq.(\ref{bgp}) (with $K=0$) and linearizing it, we write the eigenvalue equation in terms of the perturbation $\delta\psi$,
\begin{eqnarray}\label{pert1}
i\frac{\partial\delta\psi}{\partial t}&=&-\frac{1}{2}\frac{\partial^2\delta\psi}{\partial x^2}+g(2n\delta\psi+n\delta\psi^*)\nonumber\\
&&+g'(3n^{3/2}\delta\psi+n^{3/2}\delta\psi^*),
\end{eqnarray}
where $n=|\psi_0|^2$. 
A further decomposition of $\delta\psi$ in real and imaginary part leads Eq.(\ref{pert1}) to the well-known Bogoliubov-de Gennes (BdG) equation \cite{gennes}, such that,
\begin{eqnarray}\label{bdg}
&&\left(\begin{array}{c c}-\frac{1}{2}\partial_{xx}+V_1(g,g',n) & 0\\0 & -\frac{1}{2}\partial_{xx}+V_2(g,g',n)\end{array}\right)\left(\begin{array}{c}\delta\psi_R \\\delta\psi_I\end{array}\right)\nonumber\\&=&\left(\begin{array}{c c}0 & 1\\-1 & 0\end{array}\right)\partial_t\left(\begin{array}{c}\delta\psi_R \\\delta\psi_I\end{array}\right).
\end{eqnarray}
Here, $V_1(g,g',\psi_0)=gn+2g'n^{3/2}$ and $V_2(g,g',\psi_0)=3gn+4g'n^{3/2}$.
Assuming, $\delta\psi=\left(\begin{array}{c}\delta\psi_R \\\delta\psi_I\end{array}\right)=e^{i(qx-\Omega t)}$ and applying it in Eq.(\ref{bdg}), one can yield the perturbation eigenmodes where $q$ denotes the wavenumber and $\Omega$ stands for frequency. The resulting dispersion relation can be noted as $\Omega^2=(\frac{q^2}{2}+3gn+4g'n^{3/2})(\frac{q^2}{2}+gn+2g'n^{3/2})$ which boils down to
\begin{eqnarray}
\Omega^2=\frac{q^4}{4}+q^2(2gn+3g'n^{3/2}),\label{omega1}
\end{eqnarray}
by neglecting the $q$ independent terms in the dispersion relation. Further, considering the existing relationship between $g$ and $g'$ in Eq.(\ref{omega1}), we yield,
\begin{eqnarray}
\Omega^2=\frac{q^4}{4}+q^2|g|n(2-\sqrt{n}).\label{omega2}
\end{eqnarray}
\begin{figure}
\includegraphics[scale=0.5]{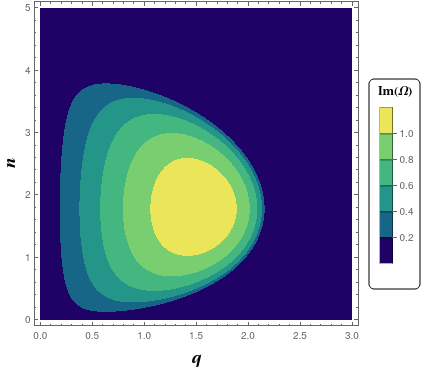}
\caption{(Color online) $\textrm{Im}(\Omega)$ or MI gain is depicted as a function of density and wavenumber using Eq.(\ref{omega2}).}\label{lin_sta}
\end{figure}
It is evident from Eq.(\ref{omega2}) that the dispersion is dependent on the sign of the right-hand side expression. If positive, then $\Omega$ will be real, corresponding to the oscillations around the unperturbed solution, however if negative, the frequency will become imaginary. This will lead to exponential growth and result instability. Therefore, one can conclude that the instability will occur when $\frac{q^4}{4}+q^2|g|n(2-\sqrt{n})<0$. Hence, if we denote $\Gamma=\textrm{Im}(\Omega)$ as MI gain then in the stable region 
$\Gamma=0$ whereas in the unstable region $\Gamma\neq 0$. Fig.\ref{lin_sta} draws the contours for $\Gamma\neq 0$ as a function of wavenumber and density. This reveals that, at low density region till $n\sim 4$ system experiences instability. From Eq.(\ref{omega2}), one can also conclude that the solution is unstable for $q^2<4|g|n(2-\sqrt{n})$.

Further, assuming $\delta\psi$ as real, we can reduce the BdG equation described in Eq.(\ref{bdg}) as,
\begin{eqnarray}\label{pert}
i\frac{\partial\delta\psi}{\partial t}=-\frac{1}{2}\frac{\partial^2\delta\psi}{\partial x^2}+3g|\psi_0|^2\delta\psi+4g'|\psi_0|^3\delta\psi.\nonumber\\
\end{eqnarray}
The perturbation equation is now effectively a Schr\"odinger equation with $\delta\psi$ being the eigenfunction. We solve the eigenvalue equation for $\delta\psi$ numerically and calculate the perturbation eigen-modes, which can be determined by considering $\delta\psi(x,t)=\delta\psi(x)e^{\Theta t}$. Here, 
\begin{eqnarray}
\Theta=\frac{\ln\{\delta\psi(x,t+\delta t)\}-\ln\{\delta\psi(x,t)\}}{\delta t}.
\end{eqnarray}
We have assumed $\delta t$ as the small numerical step length. It has been noted that, in presence of a single perturbation eigenmode, $\Theta$ will directly reduce to the corresponding eigenvalue in the limit $\delta t\rightarrow 0$ \cite{soto}. However, $\delta\psi(x)$ is expected to be composed of several perturbation eigenmodes. Albeit, for large propagation distances, the perturbation eigenmode with the largest growth rate will dominate because of the exponential nature of the growth. 
We therefore numerically check for a large propagation distance the behaviour of $\Theta$ (in the limit $\delta t\rightarrow 0$) and do not observe any exponential growth. Based on these results, from here onward, we will concentrate only on the the second solution to analyze the liquid phase.


\section{Quantum Droplet}\label{drop}
\begin{figure}
\includegraphics[scale=0.25]{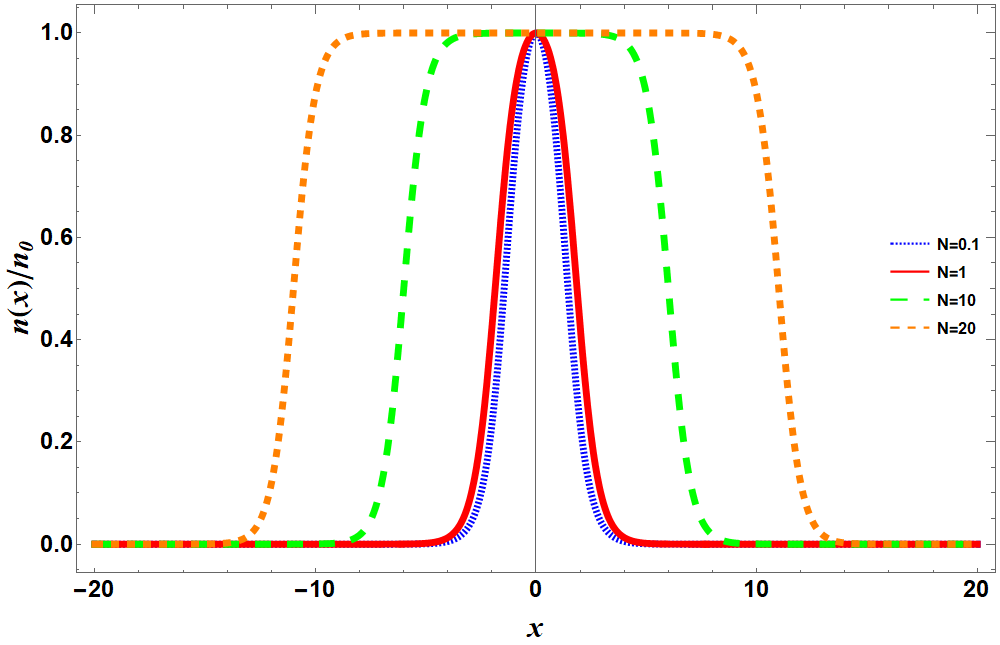}
\caption{(Color online) The stationary density profile ($n(x)=|\rho(x)|^2$) corresponding to Eq.(\ref{con2}) is depicted here. 
The blue dotted line, red solid line, green large dashed and orange short dashed lines correspond to $N=0.1, 1, 10, 20$ respectively. The density is normalized by $n_0$ where $n_0$ is $n(x)|_{x=0}$.}\label{density_plot}
\end{figure}

The signature of droplet formation can be obtained from the spatial profile of the obtained solution which we provide in Fig.~\ref{density_plot}. The figure depicts the characteristic static density profile of Eq.(\ref{con2}). However, the chemical potential $\mu_0$ is obtained by numerically solving the second equation of Eq.(\ref{N}) for different norm ($N$) at a fixed EMF ($|g|=1$). We observe a non-uniform shape for small $N$ where kinetic energy actually relevant for determining the shape as quantum pressure dominates over the potential energy. The situation is analogous to usual single component bright soliton solution with cubic nonliearity. However, as we increase $N$ we start observing a flattening of the top or accumulation of uniform density. This signature is observed for $N\geq10$ and it reminds of a classical liquid where density starts becoming spatially uniform with progressive accumulation of droplets. In the figure we have normalized all the profiles by peak density ($n_0=n(x=0)=\left|\frac{1+12\mu_g}{1+\sqrt{12\mu_g}}\right|^2$), which also happens to be the bulk value. 
It is evident from the figure that for higher $N$ the density plateau approaches the constant bulk value of $n_0$. 

It must be noted here that similar observation of density plateau is already reported in Ref.\cite{astra}. However, the density plateau was noted for a one-dimensional system which implies the governing equation was a quadratic-cubic NLSE or QCNLSE, whereas in this investigation we have concentrated on a Q1D system resulting a dynamical equation governed by cubic-quartic nonlinearities which we name CQNLSE. Another important differentiator is the nature of the nonlinearities. In the mentioned reference the quadratic and cubic nonlinearities are attractive and repulsive, respectively. In comparison, we obtain our solution for attractive cubic and repulsive quartic nonlinearity. It must be noted that the interaction strength of similar nature was involved in the experimental observation of quantum liquid in binary condensate \cite{cabrera1}.

\begin{figure}
\includegraphics[scale=0.35]{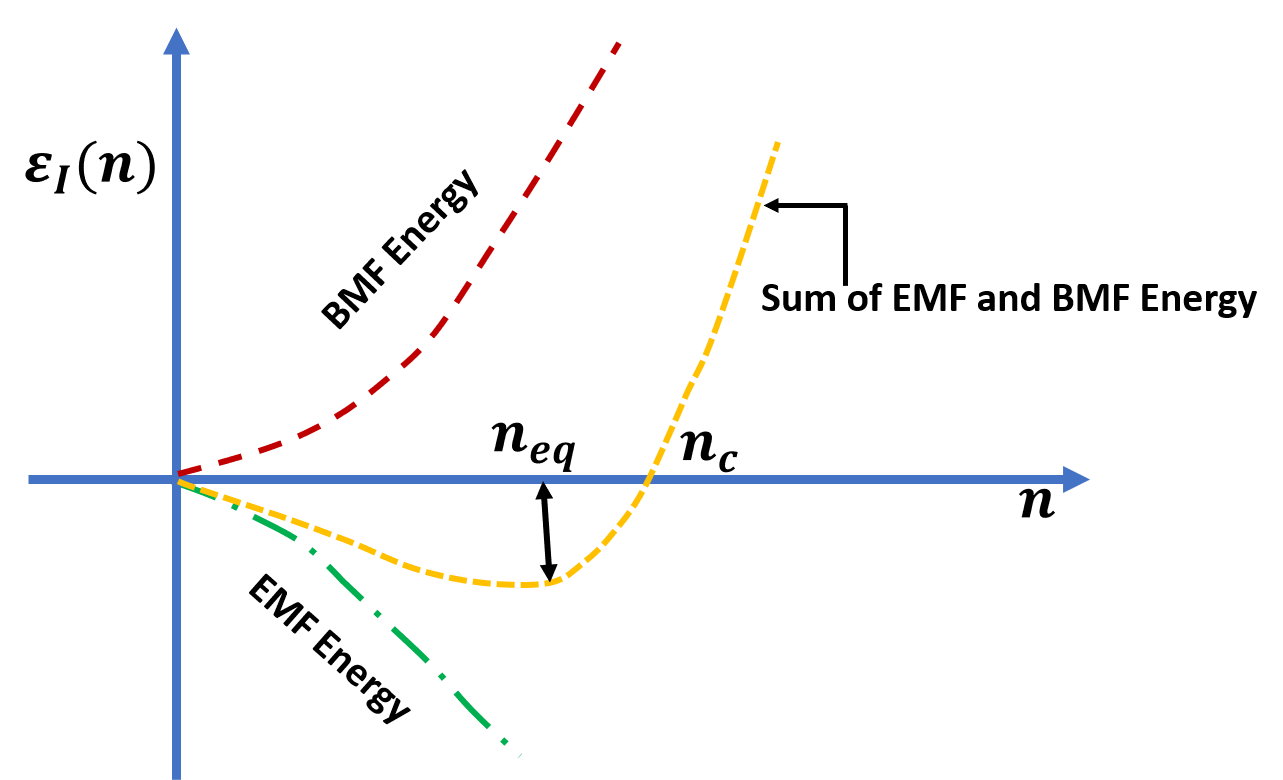}
\caption{(Color online) The attractive (EMF) and repulsive (BMF) interactions are depicted schematically. This creates an effective non-monotonic interaction. The density corresponding to the base of the effective interaction curve describes the equilibrium density ($n_{eq}$ ) and the nonzero density at which effective interaction becomes zero is defined as critical density ($n_c$).}\label{compete}
\end{figure}
\begin{figure}
\includegraphics[scale=0.25]{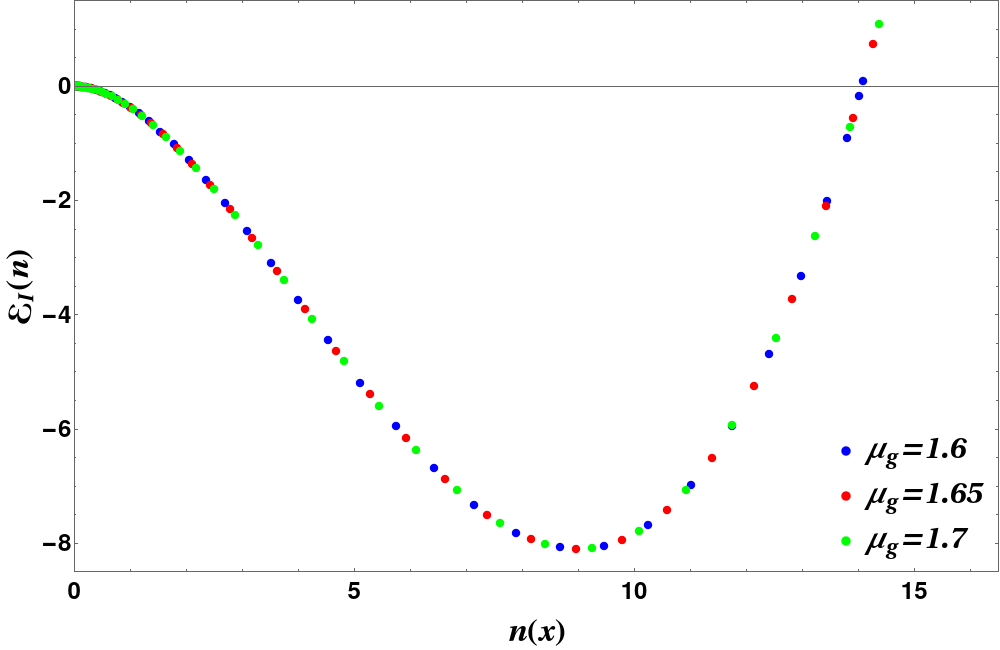}
\caption{(Color online) Interplay of effective mean-field and beyond mean-field energy resulting in droplet formation in low density using Eq.(\ref{con2}). Here, the effective mean field energy is attractive and the LHY contribution is repulsive. 
Also, we can note that the critical density is unaffected by the different $\mu_g$ values.}\label{energy_plot}
\end{figure}

As noted above, the potential energy plays the dominating role over the kinetic energy when the droplets start to accumulate and create a puddle. Thus, we concentrate on the effective potential energy which can be defined as,
$\mathcal{E}_{I}=1/2 gn^2+2/5g'n^{5/2}$. The first term is derived from the EMF interaction and the second term is BMF contribution. If both the interactions have the same sign, no
qualitative change in behavior occurs. However, the system can exhibit novel behavior if the interactions are competing as discussed in this work. When the effective mean-field interaction is low then beyond-mean-field corrections are not
necessarily negligible. Now at low density the quantum depletion remains weak, so the LHY level
approximation remains valid. Since in our case $g<0$ but $g'>0$, thus $\mathcal{E}_{I}=-1/2 gn^2+2/5g'n^{5/2}$ describes the actual effective potential. 
The situation is mimicked in Fig.~\ref{compete} where the green dashed-dotted line described the attractive EMF interaction and red dashed line depicts the repulsive BMF interaction. 
The resultant interaction is represented by a yellow short-dashed line which is initially negative at low density but slowly grows and becomes positive at relatively higher density. The level crossing point is defined as the critical density ($n_c$) after which the droplets are expected to collapse. This point can be evaluated by inserting $\mathcal{E}_I=0$ and it turns out, $n_c=\frac{25}{16}\frac{g^2}{g'^2}=14.06$ (since, $g'=|g|/3$) which matches exactly with Fig.~\ref{energy_plot}. This result also allows us to comment on the critical value in terms of interaction strength beyond which droplet formation is unlikely and it yields $\frac{\delta a}{\delta a'}<2.882$ (in units of $a_{\perp}\sqrt{a_B}$). In contrary, for a strictly one-dimensional binary system (QCNLSE), it has already been reported that the critical interaction ratio requires to be $g/|g_{12}|\leq 2.2$ \cite{astra} where $g$ and $g_{12}$ are noted as the intra and inter species coupling strength respectively. In the context of dipolar BEC the critical value was denoted as $n_{1D}a\leq 4.2$ \cite{rejish}.

The minimum of the interaction resultant signifies the equilibrium density ($n_{eq}$). At this point the pressure is zero which implies $\mathcal{P}=\mathcal{E}_I-n\frac{d\mathcal{E}_I}{dn}=0$, resulting $n_{eq}=\frac{25}{36}\frac{g^2}{g'^2}=6.25$. However, in Fig.\ref{energy_plot} the minimum is at about $8.94$. As the equilibrium density signifies the point from where the solitons start combining together to form the droplet as BMF effect takes over the EMF effect therefore to understand this anomaly is important. Hence, we analyze the chemical potential. The critical chemical potential ($\mu_{g_c}$) and equilibrium chemical potential ($\mu_{g_{eq}}$) can be expressed from effective potential energy as, $\frac{25}{64\kappa^2}=3.51$ and $-\frac{25}{216\kappa^2}=-1.04$ respectively. Using $n_c$ value from Fig.~\ref{energy_plot} if we recalculate $\mu_{g_c}$ as $\mu_{g_c}=-n_c+\kappa n_c^{3/2}$, we obtain a good agreement. The equilibrium chemical potential or $\mu_{g_{eq}}$ from Fig.~\ref{energy_plot} turns out as $-0.02$. Solving Eq.\ref{N} numerically, we also observed that when $N$ is relatively large $\mu_g\rightarrow0^{-}$ resulting the emergence of flat plateau as shown in Fig.~\ref{density_plot}. Hence, the equilibrium density obtained from Fig.~\ref{energy_plot} corroborates well with the numerical result yet the departure from the theoretical value can be attributed to the constrain condition which defines existence of exact solution only for $|g|/g'=3$. 

Nevertheless, it is well accepted that the signature of plateau is one of the important evidences of formation of the liquid-like state \cite{barbut2}. Hence, it is now possible to conclude that the droplet formation starts from the equilibrium point of density where the negative energy supports the bound state formation and further accumulation of particle happens as we increase the density till the point of $n_c$. The left-hand side of $n_{eq}$, i.e., $n<n_{eq}$, describes the bright soliton-like localized states as described in Fig.~\ref{density_plot}. 
Albeit, the system will collapse for $\mu_g=-1.5$ as solving $\mu_g=-n+\kappa n^{3/2}$ for density leads to the condition of $\mu_g=-\frac{1}{6\kappa^2}$ when the density collapses. 
One must also note here that the solution is stable in the vicinity of the liquid-like state as stability criterion leads to $-0.06\leq\mu_0\le0.0$ for unit $|g|$.

Since $\mathcal{E}_I$ has explicit dependence on density and the interaction strength, therefore a variation of $\mu_g$ does not make any significant change in $\mathcal{E}_I$ as shown in Fig.~\ref{energy_plot}.
Another noteworthy point in our calculation is that, though the liquid formation starts from negative chemical potential however, the chemical potential corresponding to critical density is positive. 
This is a notable departure from the existing understanding of quantum droplets, however, from a stability point of view, the solution is not stable for positive chemical potential. 

The observation of droplets is very recent and therefore there exists a considerable void in understanding this unique state, both theoretically and experimentally. Till now, the droplet formation in quasi-one-dimensional geometry is not yet observed. Only very recently a numerical study on droplets in quasi one dimension for dipolar BEC has been reported \cite{edmonds}. To the best of our knowledge, our analytical attempt is the very first foray in this direction albeit for binary condensate. Thus, we look forward for an experimental probing of binary condensate in quasi one dimension to validate our findings. We also plan to study the modulational instability of the obtained solution and full numerical analysis of CQNLSE in the coming days. We hope this will enrich us in understanding the phase diagram in the quasi one dimension.

\section{Conclusion}\label{con}
In this article, we study the aspect of droplet formation in a Q1D binary BEC. We start from a three-dimensional Gross Pitaevskii equation along with beyond mean-field contribution. The equation is effectively single component as we assume both the components occupy the same spatial mode \cite{cabrera2}. Then we transfer our focus towards a cigar shaped condensate for which a systematic dimensional reduction was carried out. The resulting Q1D equation is unique as it contains an additional quartic nonlinearity along with the usual cubic nonlinearity. It must be noted that there exists very little understanding of CQNLSE. Here, we derive a pair of analytical solutions corresponding to the CQNLSE. For the existence of the solution, it is required to satisfy a specific relation between the effective mean-field and beyond mean-field interaction strengths. The analytical solution is then numerically verified using split-step Crank-Nicolson method. Next, we investigate the stability of the obtained solutions using V-K criterion which results in discarding of one solution. We also study the MI to identify the physical region where the system can undergo instability. Further, we solve the eigenvalue equation pertaining to linear stability and study the behavior of the growth function against the perturbation eigenmodes.

A close inspection of the second solutions allows us to comment on the self-bound droplets as well as liquid like state. The static profile of the droplets reveals a density plateau for a higher particle number. We further analyze the effective potential energy and realize that the equilibrium chemical potential corroborates well with the numerically obtained chemical potential for flat spatial density. Moreover, the analytical solution is stable in the region where the droplets form. We also see that the critical density obtained via effective potential energy and from the analytical solution do match exactly. 

In recent description of droplet in one-dimensional geometry, the mean-field interaction is repulsive whereas the LHY term is attractive \cite{petrov1,astra} however, in our case we follow the original experimental proposition where effective two-body interaction is attractive and beyond mean-field contribution is repulsive. Our result has a close semblance with Ref.\cite{astra} however there lies a couple of fundamental differences, (i) the reported analysis is in one dimension whereas we study quasi-one-dimensional geometry. This necessitates dealing with CQNLSE instead of QCNLSE. (ii) While addressing the issue of droplet formation in a one-dimensional system \cite{astra}, the nature of the competing interaction in reverse in comparison to our model as it followed the experimental description \cite{cabrera1}. It is an undeniable fact that the topic of quantum liquid is one of the most discussed topics in the last one year or two. Therefore, we believe, our analysis of quasi-one-dimensional system will be exciting to many and will lead to the experimental verification of the current findings.

\section*{Acknowledgement} Authors acknowledge insightful discussions with T. Pfau, A. Pathak and P. Das. AK also thanks Department of Science and Technology (DST), India
for the support provided through the project number CRG/2019/000108.

\bibliographystyle{apsrev4-1}
\bibliography{ms_v4}

\end{document}